\documentclass[conference]{IEEEtran}
\IEEEoverridecommandlockouts
\usepackage[left=0.66in, right=0.66in, bottom=0.98in, top=0.75in]{geometry}

\usepackage{graphicx}
\usepackage{amsmath}
\usepackage{amsfonts}
\usepackage{amssymb}
\usepackage{algorithm}
\usepackage{algorithmic}
\usepackage{booktabs}
\ifCLASSINFOpdf

\else
\fi
\begin{document}

\title{Low-Complexity Gridless Single-Snapshot DoA Estimation via Truncated Hankel Newton-MUSIC}

\author{

\IEEEauthorblockN{Ruoxiao Cao\IEEEauthorrefmark{1}\IEEEauthorrefmark{2}, Wentao Yu\IEEEauthorrefmark{3}, Yi Gong\IEEEauthorrefmark{2}, and Khaled B. Letaief\IEEEauthorrefmark{1}}

\IEEEauthorblockA{\IEEEauthorrefmark{1}Dept. of ECE, The
    Hong Kong University of  Science and  Technology, Kowloon, Hong Kong }

\IEEEauthorblockA{\IEEEauthorrefmark{2}Dept. of EEE, Southern University of Science and Technology, Shenzhen, China}

\IEEEauthorblockA{\IEEEauthorrefmark{3}Dept. of ECE, The University of British Columbia, Vancouver, BC, V6T 1Z4, Canada }

\IEEEauthorblockA{E-mail: rcaoah@connect.ust.hk, wentaoyu@ece.ubc.ca, eekhaled@ust.hk, gongy@sustech.edu.cn}

}

\maketitle


\begin{abstract}
Reconfigurable antenna arrays can provide enhanced spatial Degrees of Freedom (DoFs) for Integrated Sensing And Communication (ISAC) systems, enabling high-resolution Direction of Arrival (DoA) estimation. In highly dynamic scenarios, however, DoA estimation must be performed within short coherence intervals, which often restricts processing to a single snapshot. Conventional subspace methods then suffer from rank deficiency, while Hankel-based spatial smoothing incurs high computational cost when the array size is large. This paper proposes a low-complexity gridless Truncated Hankel Newton-MUSIC framework for single-snapshot DoA estimation. The proposed method constructs a truncated Hankel matrix with a fixed row dimension to recover an effective signal subspace while reducing the cost of correlation construction and subspace decomposition. When the truncation length is independent of the array size, the dominant complexity scales linearly with the number of antenna ports. To reduce grid-induced quantization errors, coarse grid estimates are further refined by a second-order Newton update in the continuous angular domain. Simulation results show that the proposed method achieves DoA estimation accuracy close to square Hankel Newton-MUSIC while substantially reducing runtime. For large arrays, it provides more than two orders of magnitude runtime reduction compared with conventional square Hankel MUSIC, making it suitable for real-time sensing in reconfigurable antenna-enabled ISAC systems.
\end{abstract}

\section{Introduction}

Integrated Sensing And Communication (ISAC) is a key enabling technology for next-generation wireless networks \cite{liu2022integrated,xie2023collaborative}. By sharing spectrum resources and hardware platforms, ISAC base stations can simultaneously support high-data-rate communication and high-precision target sensing \cite{xie2025sensing}. In this context, reconfigurable antenna systems with electronically controlled radiating elements can enhance the spatial Degrees of Freedom (DoFs) available for sensing \cite{wong2020fluid,zhu2023movable}. Reconfigurable antenna-enabled ISAC is particularly attractive because the spatial sampling pattern can be adapted through port selection or element reconfiguration, providing additional flexibility beyond conventional fixed arrays \cite{yu2026sensing,cao2026sensing}. Such flexibility can improve angular resolution and sensing robustness without requiring fully digital radio-frequency chains at all antenna ports. However, the increased number of available ports also enlarges the dimension of the sensing data, making low-latency Direction of Arrival (DoA) estimation a critical signal processing challenge.

Despite these advantages, DoA estimation remains challenging in highly dynamic mobile environments. To enable real-time tracking in high-mobility scenarios, parameter estimation often needs to be completed within a short coherence interval, which may restrict processing to a single snapshot \cite{krim1996two}. Conventional high-resolution subspace methods, such as MUltiple SIgnal Classification (MUSIC) \cite{schmidt1986multiple} and Estimation of Signal Parameters via Rotational Invariance Techniques (ESPRIT) \cite{roy1989esprit}, rely on temporal averaging over multiple independent snapshots to construct a full-rank sample covariance matrix. When applied to a single snapshot, the sample covariance matrix becomes rank-deficient and collapses to rank one, making the signal and noise subspaces difficult to separate \cite{krim1996two,schmidt1986multiple}. This prevents the separation of multiple signal components and leads to the failure of subspace-based resolution for closely spaced targets.

Spatial smoothing is a common approach to mitigate single-snapshot rank deficiency \cite{shan1985spatial}. By rearranging one array snapshot into a Hankel-structured matrix, multiple overlapping subarrays can be generated from spatial samples, enabling subspace recovery without temporal averaging \cite{liao2016music,fannjiang2025optimality}. In many Hankel-MUSIC implementations, the smoothing window is chosen close to half of the array aperture to balance the row and column dimensions of the Hankel matrix. While this choice is effective for subspace restoration, it results in a large smoothed correlation matrix. The corresponding correlation construction and subspace decomposition become computationally expensive for large reconfigurable arrays.

Another limitation of conventional MUSIC-based DoA estimation is the dependence on discrete angular grid search. A dense grid is needed to reduce quantization error, but the search complexity increases linearly with the number of grid points \cite{fannjiang2025optimality}. This tradeoff is undesirable for real-time ISAC sensing, where both low latency and high angular accuracy are required. Off-grid refinement can alleviate this issue by updating coarse estimates in the continuous angular domain \cite{mamandipoor2016newtonized,cao2023joint}, but it must be combined with a low-dimensional subspace representation to be computationally efficient.

To address these limitations, this paper proposes a low-complexity truncated Hankel Newton-MUSIC framework for ISAC systems. The proposed method leverages the sparsity of propagation paths in the angular domain and applies structural truncation by restricting the row dimension of the Hankel smoothing operation to a small constant. This reformulates the single snapshot into a rectangular Hankel structure, resulting in a compact low-dimensional representation while preserving the effective array aperture through its extended column dimension. Although truncation reduces the spatial DoFs, the remaining structure preserves sufficient spatial diversity for effective smoothing. To compensate for the resulting resolution loss, a second-order Newton refinement scheme is introduced to iteratively refine coarse estimates toward continuous-valued directions. Simulation results demonstrate that the proposed framework reduces computational complexity by more than two orders of magnitude compared with conventional square Hankel methods, while mitigating the high-SNR error floor caused by grid-induced quantization errors.

\emph{Notations:} For any matrix $\mathbf{A}$, $\mathbf{A}^{\textrm{T}}$,
$\mathbf{A}^{-1}$ and $\mathbf{A}^{\textrm{H}}$ denote the transpose, inverse and conjugate transpose of $\mathbf{A}$, respectively. $\mathbf{I}_{N}$ represents the identity matrix of
size $N\times N$. The Euclidean norm is written as $\|\cdot\|_{2}$.
$\mathcal{CN}(\boldsymbol{\mu},\mathbf{\Sigma})$ denotes the complex
Gaussian distribution with mean vector $\boldsymbol{\mu}$ and covariance
matrix $\mathbf{\Sigma}$. $\mathcal{U}\left(a,b\right)$ denotes
the uniform distribution from $a$ to $b$. The operators $\operatorname{diag}(\cdot)$ and $\Re\{\cdot\}$ generate a diagonal matrix and extract the real part, respectively. The imaginary unit is $j$.

\section{System Model and Problem Formulation}
In this section, we present the signal model for single-snapshot DoA estimation in a reconfigurable antenna system. To ensure geometric robustness and avoid specific hardware non-idealities, we consider that during the initial stage of directional sensing, the reconfigurable antenna array is dynamically configured into a standard Uniform Linear Array (ULA) layout, which provides optimized aperture geometry to achieve enhanced angular estimation precision.

\subsection{Spatial Signal Model}
Assume that $K$ narrowband far-field sources transmit signals simultaneously to the ISAC base station (BS) in the uplink. Under the constraint of a single temporal snapshot ($T=1$) during a dedicated pilot transmission phase, let $\mathbf{s} = [s_1, s_2, \dots, s_K]^{\textrm{T}} \in \mathbb{C}^{K \times 1}$ denote the transmitted pilot signal vector, where $s_k$ represents the deterministic and known pilot symbol of the $k$-th source. To model the angular characteristics of the propagation environment, the received signal vector captured at the $N$ activated ports of the reconfigurable antenna system, denoted by $\mathbf{y} \in \mathbb{C}^{N \times 1}$, is expressed as
\begin{equation}
\mathbf{y} = \mathbf{A}(\boldsymbol{\theta})\mathbf{P}\mathbf{s} + \mathbf{n}, \label{eq:SignalModel}
\end{equation}
where $\mathbf{A}(\boldsymbol{\theta}) = [\mathbf{a}(\theta_1), \mathbf{a}(\theta_2), \dots, \mathbf{a}(\theta_K)] \in \mathbb{C}^{N \times K}$ denotes the array response matrix, $\mathbf{P} = \text{diag}(p_1, p_2, \dots, p_K) \in \mathbb{C}^{K \times K}$ is a diagonal matrix containing the complex path gains, and $\mathbf{n} \in \mathbb{C}^{N \times 1}$ is the complex additive white Gaussian noise (AWGN) vector distributed as $\mathbf{n} \sim \mathcal{CN}(\mathbf{0}, \sigma^2 \mathbf{I}_N)$. The variance of the noise across the spatial ports is denoted by $\sigma^2$, and the vector $\boldsymbol{\theta} = [\theta_1, \theta_2, \dots, \theta_K]^{\textrm{T}}$ contains the DoAs of the target paths.

Let $\mathbf{d} = [d_1, d_2, \dots, d_{N}]^{\textrm{T}} \in \mathbb{R}^{N \times 1}$ denote the physical spatial positions of the activated ports in the reconfigurable antenna array. To optimize the spatial resolution and establish a well-defined sensing geometry, the reconfigurable antenna system configures its electronic port components into a linear topology during the initial DoA estimation stage. By setting the array into this baseline geometry during the spatial sensing slot, the physical inter-port displacement satisfies $d_\textrm{n} - d_1 = (n-1)\frac{\lambda}{2}$ for $n \in \{1, 2, \dots, N\}$, where $\lambda$ represents the carrier wavelength. Under this configuration, the generic geometric steering response vector $\mathbf{a}(\theta_k)$ matches the standard nonharmonic exponential manifold form given by
\begin{equation}
\mathbf{a}(\theta_k) = \left[1, e^{-j \pi \sin(\theta_k)}, e^{-j \pi 2 \sin(\theta_k)}, \dots, e^{-j \pi (N-1) \sin(\theta_k)}\right]^{\textrm{T}}.
\end{equation}
The formulated single-snapshot observation relation in \eqref{eq:SignalModel} directly serves as the mathematical foundation for subsequent low-complexity subspace extractions over the reconfigurable antenna architecture.

\subsection{The Single-Snapshot Rank-Deficiency Dilemma}
In traditional multi-snapshot subspace estimation architectures, the receiver accumulates a large temporal sequence of observation vectors to form a statistical sample covariance matrix. When the source signals are temporally uncorrelated, the empirical covariance matrix asymptotically approaches a full-rank signal subspace of rank $K$ to enable separation between the signal and noise subspaces.

Conversely, in highly dynamic networks where online state optimization within the reconfigurable antenna system must be completed within a short coherence interval, the receiver is restricted to the single-snapshot vector $\mathbf{y}$ defined in (1). If the raw spatial correlation matrix $\mathbf{R}_{\text{raw}} \in \mathbb{C}^{N \times N}$ is constructed via direct outer-product mapping, it is given by
\begin{equation}
\mathbf{R}_{\text{raw}} = \mathbf{y}\mathbf{y}^\textrm{H} = \left( \mathbf{A}(\boldsymbol{\theta})\mathbf{P}\mathbf{s} + \mathbf{n} \right)\left( \mathbf{A}(\boldsymbol{\theta})\mathbf{P}\mathbf{s} + \mathbf{n} \right)^\textrm{H},
\end{equation}
where rank deficiency arises due to the lack of temporal averaging. By invoking the rank properties of matrix outer products, the algebraic rank of the noise-free component of $\mathbf{R}_{\text{raw}}$ collapses completely as expressed by
\begin{equation}
\text{rank}\left(\mathbf{A}(\boldsymbol{\theta})\mathbf{P}\mathbf{s}\mathbf{s}^\textrm{H}\mathbf{P}^\textrm{H}\mathbf{A}^\textrm{H}(\boldsymbol{\theta})\right) = \text{rank}(\mathbf{P}\mathbf{s}\mathbf{s}^\textrm{H}\mathbf{P}^\textrm{H}) \equiv 1.
\end{equation}

This rank deflation implies that the true multidimensional signal subspace contracts into a single-dimensional line, rendering the remaining $K-1$ signal components indistinguishable from the background thermal noise. As a direct consequence, classical high-resolution subspace algorithms fail to resolve the individual peaks of closely spaced targets. To address the rank deficiency caused by the single-snapshot observation $\mathbf{y}$, we construct a Hankel-structured matrix from its spatial samples to recover an effective signal subspace for subsequent MUSIC processing.

\section{Proposed Low-Complexity Newton-MUSIC Framework}
To address the rank deficiency caused by single-snapshot sensing, we develop a low-complexity gridless Newton-MUSIC framework for reconfigurable antenna-enabled ISAC systems. We first formulate a general Hankel-based single-snapshot MUSIC method to recover the effective signal subspace from spatially smoothed observations. We then introduce a truncated Hankel construction that reduces the row dimension of the smoothed data matrix while retaining a large number of overlapping subarray snapshots through its column dimension. Finally, a Newton refinement step is applied to improve the coarse grid estimates in the continuous angular domain and reduce grid-induced quantization errors.

\subsection{General Single-Snapshot MUSIC Framework}
To recover the effective rank from a single snapshot, spatial smoothing constructs overlapping subarrays from the received array snapshot. For a generic smoothing window length $L$ satisfying $1 \le L < N$, the received observation vector $\mathbf{y}$ is mapped into a general structured Hankel data matrix $\mathbf{H}_L \in \mathbb{C}^{(L+1) \times (N-L)}$ which is written as
\begin{equation}
\mathbf{H}_L = \textrm{Hankel}\left(\mathbf{y}\right) = \begin{bmatrix} y_1 & y_2 & \cdots & y_{N-L} \\ y_2 & y_3 & \cdots & y_{N-L+1} \\ \vdots & \vdots & \ddots & \vdots \\ y_{L+1} & y_{L+2} & \cdots & y_{N} \end{bmatrix}.
\label{eq:HankelMatrix}
\end{equation}
where $y_i$ is the $i$th element of the received observation vector $\mathbf{y}$. The corresponding spatial correlation matrix $\mathbf{R} \in \mathbb{C}^{(L+1) \times (L+1)}$ is computed via the inner product mapping
\begin{equation}
\mathbf{R}_L = \frac{1}{N-L} \mathbf{H}_L\mathbf{H}_L^\textrm{H}. \label{eq:CorrelationMatrix}
\end{equation}
Performing singular value decomposition (SVD) on $\mathbf{R}_L$ decomposes the matrix as\footnote{Performing the SVD on the spatial correlation matrix $\mathbf{R}_L$ is mathematically consistent with executing the SVD directly on the rectangular Hankel data matrix $\mathbf{H} = \mathbf{U}_H \boldsymbol{\Sigma}_H \mathbf{V}_H^\textrm{H}$. Since $\mathbf{R}_L$ is formed by the scaled outer product of $\mathbf{H}_L$, the left singular vectors $\mathbf{U}$ obtained from the SVD of $\mathbf{R}_L$ align with the left singular vectors $\mathbf{U}_H$ obtained from the direct SVD of $\mathbf{H}_L$. The corresponding singular values satisfy the quadratic relation $\boldsymbol{\Sigma} = \frac{1}{N-L}\boldsymbol{\Sigma}_H^2$. This algebraic equivalence guarantees that both formulations yield identical orthogonal signal and noise subspaces while performing SVD on $\mathbf{R}_L$ preserves the conventional covariance-driven analytical paradigm.}
\begin{equation}
\begin{split}
\mathbf{R} &= \mathbf{U} \boldsymbol{\Sigma} \mathbf{V}^\textrm{H} \\
&= \begin{bmatrix} \mathbf{U}_\textrm{s} & \mathbf{U}_\textrm{n} \end{bmatrix} \begin{bmatrix} \boldsymbol{\Sigma}_\textrm{s} & \mathbf{0} \\ \mathbf{0} & \boldsymbol{\Sigma}_\textrm{n} \end{bmatrix} \begin{bmatrix} \mathbf{V}_\textrm{s}^\textrm{H} \\ \mathbf{V}_\textrm{n}^\textrm{H} \end{bmatrix},
\end{split} \label{eq:SVD}
\end{equation}
where $\mathbf{U}_\textrm{s} \in \mathbb{C}^{(L+1) \times K}$ and $\mathbf{V}_\textrm{s} \in \mathbb{C}^{(L+1) \times K}$ contain the left and right singular vectors corresponding to the $K$ largest singular values, which span the signal subspace. The matrices $\mathbf{U}_\textrm{n} \in \mathbb{C}^{(L+1) \times (L+1-K)}$ and $\mathbf{V}_\textrm{n} \in \mathbb{C}^{(L+1) \times (L+1-K)}$ contain the remaining singular vectors spanning the orthogonal noise subspace. The diagonal matrices $\boldsymbol{\Sigma}_\textrm{s} \in \mathbb{R}^{K \times K}$ and $\boldsymbol{\Sigma}_\textrm{n} \in \mathbb{R}^{(L+1-K) \times (L+1-K)}$ contain the corresponding singular values sorted in descending order. The general continuous spatial spectrum function is defined as the reciprocal of the geometric projection of the steering vector onto the noise subspace, expressed as
\begin{equation}
F(\theta) = \frac{1}{\mathbf{a}_{L}^\textrm{H}(\theta)\mathbf{U}_\textrm{n}\mathbf{U}_\textrm{n}^\textrm{H}\mathbf{a}_{L}(\theta)},
\end{equation}
where $\mathbf{a}_{L}(\theta) \in \mathbb{C}^{(L+1) \times 1} $ represents the low-dimensional steering response vector configured under the truncated sub-array topology, which is explicitly given by
\begin{equation}
\mathbf{a}_{L}(\theta) = \left[1, e^{-j \pi \sin(\theta)}, e^{-j \pi 2 \sin(\theta)}, \dots, e^{-j \pi L \sin(\theta)}\right]^\textrm{T}. \label{eq:LSteeringVector}
\end{equation} 
MUSIC exploits the orthogonality between the steering vectors associated with the true DoAs and the estimated noise subspace. Since the signal subspace is spanned by the columns of the array response matrix $\mathbf{A}_L(\boldsymbol{\theta}) = [\mathbf{a}_L(\theta_1), \mathbf{a}_L(\theta_2), \dots, \mathbf{a}_L(\theta_K)] \in \mathbb{C}^{(L+1) \times K}$, any true steering vector must be orthogonal to the noise subspace, i.e., $\mathbf{a}_{L}^\textrm{H}(\theta)\mathbf{U}_\textrm{n}=0$. In the noiseless and perfectly matched case, the denominator of $F(\theta)$ vanishes at the true DoAs, producing peaks in the MUSIC spectrum.  The DoA estimates are subsequently identified by locating the $K$ largest local peaks of the spatial spectrum over a discrete angular domain $\Theta$ as given by
\begin{equation}
\hat{\theta}_k = \arg\max_{\theta \in \Theta} F(\theta), \label{eq:GridSearch}
\end{equation}
where $\hat{\boldsymbol{\theta}}=\left\{ \hat{\theta}_{k}:k=1,2,\ldots,K\right\}$  is the estimated DoAs.

In conventional implementations, the smoothing window parameter is typically fixed at approximately half the total port aperture such that $L_\textrm{O} \approx N/2$ \cite{liao2016music, fannjiang2025optimality}. This configuration is referred to as \textit{square Hankel MUSIC} in this paper. While this symmetric square configuration maximizes the available spatial DoFs and stabilizes the subspace boundaries against noise perturbations, computing a full SVD on a dense $N/2 \times N/2$ matrix introduces a prohibitive cubic computational overhead of $\mathcal{O}(N^3)$ complexity. In addition, the conventional discrete spectrum peak-search introduces inherent on-grid quantization errors, which can only be mitigated by employing a significantly denser angular search grid $\Theta$ that consequently intensifies the grid-search computational burden.

\subsection{Proposed Single-Snapshot Newton-MUSIC Algorithm}
To reduce the cost of high-dimensional subspace decomposition and alleviate grid mismatch, we propose a truncated Hankel Newton-MUSIC framework. The method combines a low-dimensional Hankel smoothing structure with a continuous Newton refinement step, avoiding the need for a dense angular search grid.

Since the number of sources is small compared with the array size, i.e., $K\ll N$, it is unnecessary to use a nearly square Hankel matrix for subspace recovery. We therefore restrict the row dimension of the Hankel matrix to a small value $L_{\mathrm{N}}$ satisfying $L_{\mathrm{N}}+1>K$ and $L_{\mathrm{N}}\ll N$. This choice preserves a nonempty noise subspace of dimension $L_{\mathrm{N}}+1-K$. Although the row dimension is reduced, the column dimension $N-L_{\mathrm{N}}$ still provides many overlapping subarrays, which improves the stability of the smoothed correlation estimate. The possible loss in resolution caused by truncation is mitigated by the subsequent Newton refinement in the continuous angular domain.

To convert the discrete search grid in \eqref{eq:GridSearch} into a continuous optimization space, we define the MUSIC projection cost as
\begin{equation}
J(\theta) = \mathbf{a}_L^\textrm{H}(\theta)\mathbf{U}_\textrm{n}\mathbf{U}_\textrm{n}^\textrm{H}\mathbf{a}_L(\theta).
\end{equation}
The DoA estimates can therefore be obtained by identifying the $K$ local minima of $J(\theta)$, which is formulated as
\begin{equation}
\hat{\theta}_k = \arg\min_{\theta} J(\theta).
\end{equation}
We employ a coarse grid search with large angular spacing $\Theta_{\text{coarse}}$ to locate the initial rough estimates denoted by $\hat{\boldsymbol{\theta}}^{(0)}=\left\{ \hat{\theta}_{k}^{(0)}:k=1,2,\ldots,K\right\}$. Given these initial values, the Newton refinement updates the parameter estimates iteratively through the second-order optimization relationship
\begin{equation}
\hat{\theta}_k^{(z)} = \hat{\theta}_k^{(z-1)} - \frac{ \nabla J\!\left(\hat{\theta}_k^{(z-1)}\right) }{ \nabla^2 J\!\left(\hat{\theta}_k^{(z-1)}\right) }, \label{eq:NewtonRefinement} 
\end{equation}
where $\nabla J(\theta)$ and $\nabla^2 J(\theta)$ denote the gradient and the Hessian of the cost function with respect to $\theta$, respectively, and $z \in \{1, 2, \dots, Z\}$ represents the local iteration index.

\begin{algorithm}[t]
\caption{Proposed Single-Snapshot Newton-MUSIC}
\label{alg:newton_music}
\begin{algorithmic}[1]
\renewcommand{\algorithmicrequire}{\textbf{Input:}}
\renewcommand{\algorithmicensure}{\textbf{Output:}}
\REQUIRE Received single snapshot vector $\mathbf{y}$, number of target sources $K$, coarse angular grid $\Theta_{\text{coarse}}$, maximum number of iterations $Z$, truncation window length $L_\textrm{N}$.
\STATE Construct the truncated Hankel matrix $\mathbf{H}_{L}$ from $\mathbf{y}$ according to \eqref{eq:HankelMatrix}.
\STATE Compute the smoothed correlation matrix $\mathbf{R}_{L}$ using \eqref{eq:CorrelationMatrix}.
\STATE Perform SVD on $\mathbf{R}_{L}$ and obtain the noise subspace $\mathbf{U}_{\mathrm{n}}$ via \eqref{eq:SVD};
\STATE Evaluate $J(\theta)$ on $\Theta_{\mathrm{coarse}}$ and select the $K$ smallest local minima as the initial estimates $\{\hat{\theta}_k^{(0)}\}_{k=1}^{K}$.
\FOR{$k = 1$ \TO $K$ \textbf{in parallel}}
    \FOR{$z = 1$ \TO $Z$}
        \STATE Compute $\mathbf{a}_{L}(\hat{\theta}_k^{(z-1)})$, $\dot{\mathbf{a}}_{L}(\hat{\theta}_k^{(z-1)})$, and $\ddot{\mathbf{a}}_{L}(\hat{\theta}_k^{(z-1)})$.   
        \STATE Evaluate $\nabla J(\hat{\theta}_k^{(z-1)})$ and $\nabla^2 J(\hat{\theta}_k^{(z-1)})$ using \eqref{eq:Gradient} and \eqref{eq:Hessian}.
        \STATE Update $\hat{\theta}_k^{(z)}$ using \eqref{eq:NewtonRefinement}.
    \ENDFOR
    \STATE Set $\hat{\theta}_k=\hat{\theta}_k^{(z)}$.
\ENDFOR
\ENSURE Estimated off-grid DoAs $\hat{\boldsymbol{\theta}} = [\hat{\theta}_1, \hat{\theta}_2, \dots, \hat{\theta}_K]^\textrm{T}$.
\end{algorithmic}
\end{algorithm}

To calculate these optimization components analytically, the first and second derivatives of the steering manifold are derived. Applying the chain rule to the complex exponential elements yields the precise analytical derivative vectors given by
\begin{equation}
\dot{\mathbf{a}}_{L}(\theta) \triangleq \frac{\partial \mathbf{a}_{L}(\theta)}{\partial \theta} = -j \pi \cos(\theta) \mathbf{D}_{L} \mathbf{a}_{L}(\theta), \label{eq:1stDerivative}
\end{equation}
\begin{equation}
    \ddot{\mathbf{a}}_{L}(\theta) \triangleq \frac{\partial^2 \mathbf{a}_{L}(\theta)}{\partial \theta^2} = \left( j \pi \sin(\theta) \mathbf{D}_{L} - \pi^2 \cos^2(\theta) \mathbf{D}_{L}^2 \right) \mathbf{a}_{L}(\theta), \label{eq:2ndDerivative}
\end{equation}
where $\mathbf{D}_{L} = \text{diag}(0, 1, \dots, L_\textrm{N}) \in \mathbb{R}^{(L_\textrm{N}+1) \times (L_\textrm{N}+1)}$ represents the structural diagonal position matrix. By utilizing these matrix derivative vectors, the exact gradient function of the continuous cost objective is formulated as
\begin{equation}
\nabla J(\theta) = 2 \Re \left\{ \dot{\mathbf{a}}_L^\textrm{H}(\theta) \mathbf{U}_\textrm{n} \mathbf{U}_\textrm{n}^\textrm{H} \mathbf{a}_L(\theta) \right\}, \label{eq:Gradient}
\end{equation}
and the corresponding analytical Hessian expression is established as
\begin{equation}
\nabla^2 J(\theta) = 2  \Re \left\{ \dot{\mathbf{a}}_L^\textrm{H}(\theta) \mathbf{U}_\textrm{n} \mathbf{U}_\textrm{n}^\textrm{H} \dot{\mathbf{a}}_L(\theta) + \mathbf{a}_L^\textrm{H}(\theta) \mathbf{U}_\textrm{n} \mathbf{U}_\textrm{n}^\textrm{H} \ddot{\mathbf{a}}_L(\theta) \right\} . \label{eq:Hessian}
\end{equation}
The continuous parameter tracking updates terminate after a maximum number of local iterations $Z$.

The complete procedure of the proposed truncated Hankel Newton-MUSIC algorithm is summarized in Algorithm \ref{alg:newton_music}. Compared with conventional single-snapshot MUSIC, the proposed method avoids exhaustive search over a dense angular grid by refining coarse estimates through local Newton updates. This reduces the DoA extraction cost while retaining off-grid estimation capability. After the coarse initialization, each Newton refinement is performed independently for its corresponding local minimum, and the $K$ refinement processes can therefore be parallelized.


\begin{table*}[t]
\centering
\caption{Computational Complexity Comparison}
\renewcommand{\arraystretch}{1.3}
\small
\begin{tabular}{lll}
\toprule
\textbf{Processing Stage} & \textbf{Traditional Square Hankel MUSIC} & \textbf{Proposed Truncated Hankel Newton-MUSIC} \\ \midrule
Correlation Construction  & $\mathcal{O}\!\left(L_{\mathrm{O}}^{2}(N-L_{\mathrm{O}})\right)\approx \mathcal{O}(N^3)$  & $\mathcal{O}\!\left(L_{\mathrm{N}}^{2}(N-L_{\mathrm{N}})\right)\approx \mathcal{O}(L_{\mathrm{N}}^{2}N)$ \\
Subspace Decomposition  & $\mathcal{O}\!\left((L_{\mathrm{O}}+1)^3\right)\approx \mathcal{O}(N^3)$  & $\mathcal{O}\!\left((L_{\mathrm{N}}+1)^3\right)\approx \mathcal{O}(L_{\mathrm{N}}^3)$ \\
DoA Extraction  & $\mathcal{O}\!\left(|\Theta|L_{\mathrm{O}}^2\right)\approx \mathcal{O}(|\Theta|N^2)$  & $\mathcal{O}\!\left(|\Theta_{\mathrm{coarse}}|L_{\mathrm{N}}^2+ZL_{\mathrm{N}}^2\right)$ \\ \midrule
Overall Complexity  & $\mathcal{O}\!\left(N^3+|\Theta|N^2\right)$  & $\mathcal{O}\!\left(L_{\mathrm{N}}^2N+L_{\mathrm{N}}^3+|\Theta_{\mathrm{coarse}}|L_{\mathrm{N}}^2+ZL_{\mathrm{N}}^2\right)$ \\
\bottomrule 
\end{tabular}
\label{tab:ComplexityComparison}
\end{table*}

\section{Computational Complexity Analysis}
The computational cost is analyzed from two aspects: Hankel-based subspace construction and DoA extraction. The first part includes Hankel matrix formation, correlation construction, and subspace decomposition, while the second part accounts for grid-based spectrum evaluation and Newton refinement. Table \ref{tab:ComplexityComparison} compares the conventional square Hankel MUSIC method with the proposed truncated Hankel Newton-MUSIC method. The reduction in complexity comes from two design choices: using a truncated Hankel matrix with a fixed row dimension and replacing the dense grid search with local Newton refinement initialized from a coarse grid.

\subsection{Hankel Dimension Reduction and Subspace Decomposition}

In square Hankel MUSIC, the smoothing window is usually chosen as $L_{\mathrm{O}}\approx N/2$ to balance the row and column dimensions of the Hankel matrix. The resulting correlation matrix has dimension $(L_{\mathrm{O}}+1)\times(L_{\mathrm{O}}+1)$. With $L_{\mathrm{O}}\approx N/2$, constructing $\mathbf{R}_{L_{\mathrm{O}}}=\mathbf{H}_{L_{\mathrm{O}}}\mathbf{H}_{L_{\mathrm{O}}}^{\mathrm{H}}$ requires $\mathcal{O}(L_{\mathrm{O}}^2(N-L_{\mathrm{O}}))$, which scales as $\mathcal{O}(N^3)$. The subsequent SVD of $\mathbf{R}_{L_{\mathrm{O}}}$ requires $\mathcal{O}((L_{\mathrm{O}}+1)^3)$ operations, which also scales as $\mathcal{O}(N^3)$.

In the proposed method, the row dimension is fixed to a small truncation length $L_{\mathrm{N}}$ satisfying $L_{\mathrm{N}}\ll N$. The smoothed correlation matrix then has dimension $(L_{\mathrm{N}}+1)\times(L_{\mathrm{N}}+1)$. Its construction requires $\mathcal{O}(L_{\mathrm{N}}^2(N-L_{\mathrm{N}}))$, which reduces to $\mathcal{O}(L_{\mathrm{N}}^2N)$ for $L_{\mathrm{N}}\ll N$. The SVD requires only $\mathcal{O}((L_{\mathrm{N}}+1)^3)$ operations. Therefore, when $L_{\mathrm{N}}$ is fixed independently of $N$, the dominant complexity of the proposed method scales linearly with the array size.

\subsection{DoA Extraction and Newton Refinement}

The DoA extraction stage further reduces the search cost by replacing a dense grid search with a coarse initialization followed by Newton refinement. Conventional MUSIC evaluates the spatial spectrum over a dense angular grid $\Theta$. Evaluating the MUSIC cost over this grid requires $\mathcal{O}(|\Theta|L_{\mathrm{O}}^2)$ operations when the noise-subspace projection is directly used. A finer grid reduces quantization error but increases the search cost proportionally to $|\Theta|$.

The proposed method uses a two-stage estimation procedure. First, the cost function is evaluated on a coarse grid $\Theta_{\mathrm{coarse}}$, which requires $\mathcal{O}(|\Theta_{\mathrm{coarse}}|L_{\mathrm{N}}^2)$ operations. Since $|\Theta_{\mathrm{coarse}}|\ll|\Theta|$, this step is much cheaper than dense grid search. Second, each coarse estimate is refined by $Z$ Newton iterations in the continuous angular domain. The total arithmetic cost of the refinement stage is $\mathcal{O}(KZL_{\mathrm{N}}^2)$. After coarse initialization, each local Newton refinement can be carried out independently, so the $K$ refinement loops can be parallelized. Under a parallel implementation, the refinement latency scales as $\mathcal{O}(ZL_{\mathrm{N}}^2)$.

\section{Simulation Results}
We evaluate the proposed method using Monte Carlo simulations. Unless otherwise specified, the number of antenna ports is set to $N=256$, and the number of far-field sources is $K=4$. The source DoAs are randomly generated within $\theta\in[-60^\circ,60^\circ]$ with a minimum angular separation of $\Delta\theta=9^\circ$. The smoothing lengths are set to $L_{\mathrm{O}}=128$ for square Hankel methods and $L_{\mathrm{N}}=20$ for truncated Hankel methods. Each data point is averaged over $M=2000$ independent Monte Carlo trials. The path powers are generated independently as $p_k\sim\mathcal{U}(p_{\min},p_{\max})$, where $p_{\max}/p_{\min}\leq 10$. The average received Signal-to-Noise Ratio (SNR) is defined as
\begin{equation}
\mathrm{SNR}\,[\mathrm{dB}]
\triangleq
10\log_{10}\left(
\frac{\frac{1}{K}\sum_{k=1}^{K}p_k}{\sigma^{2}}
\right).
\label{eq:SNR}
\end{equation}
For grid-search MUSIC, the dense angular grid is defined as
\begin{equation}
\Theta=\{\theta_{\min}:\Delta_{\Theta}:\theta_{\max}\},
\end{equation}
with grid spacing $\Delta_{\Theta}=0.1^\circ$. For Newton-MUSIC, the coarse grid is defined as
\begin{equation}
\Theta_{\mathrm{coarse}}
=
\{\theta_{\min}:\Delta_{\Theta_{\mathrm{coarse}}}:\theta_{\max}\},
\end{equation}
with $\Delta_{\Theta_{\mathrm{coarse}}}=0.5^\circ$. The maximum number of Newton iterations is set to $Z=20$. 

To evaluate the individual effects of Hankel truncation and Newton refinement, we compare four algorithmic configurations:
\begin{itemize}
    \item \textbf{Square Hankel MUSIC}: the conventional Hankel-MUSIC method using a square Hankel matrix and dense angular grid search \cite{liao2016music}.
    \item \textbf{Truncated Hankel MUSIC}: a truncated Hankel variant using the reduced row dimension $L_{\mathrm{N}}$ while still performing dense angular grid search.
    \item \textbf{Square Hankel Newton-MUSIC}: a square Hankel variant that applies Newton refinement to coarse grid estimates.
    \item \textbf{Truncated Hankel Newton-MUSIC}: the proposed method, which combines truncated Hankel smoothing with Newton refinement in the continuous angular domain.
\end{itemize}

\begin{figure}[t]
\centering
\includegraphics[width=0.88\linewidth]{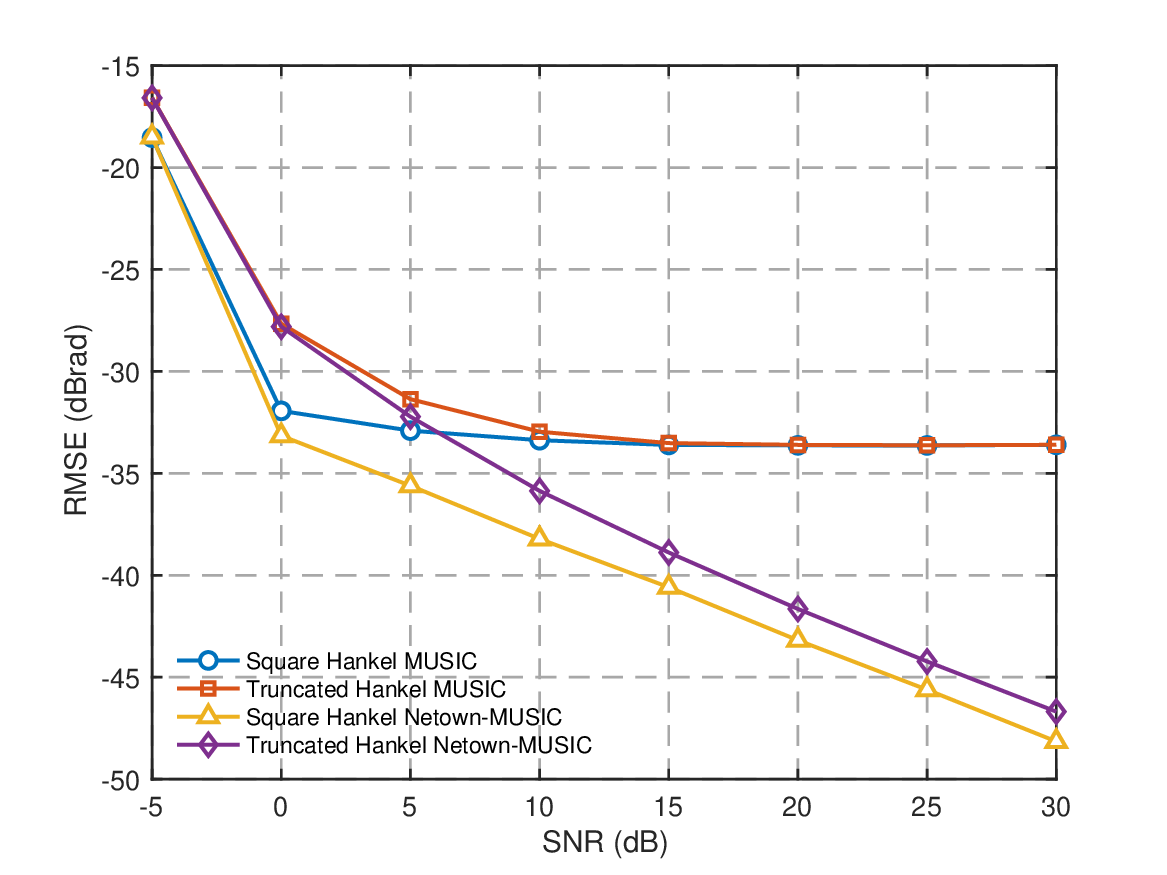}
\caption{RMSE performance of DoA estimation versus SNR.}
\label{fig:rmse}
\end{figure}

The Root-Mean-Square Error (RMSE) of the DoA estimates is defined as
\begin{equation}
\operatorname{RMSE}
=
\sqrt{
\frac{1}{MK}
\sum_{m=1}^{M}
\min_{\mathbf{\Pi}\in\mathcal{P}_{K}}
\left\|
\mathbf{\Pi}\hat{\boldsymbol{\theta}}^{(m)}
-
\boldsymbol{\theta}^{(m)}
\right\|_2^2
},
\label{eq:RMSE}
\end{equation}
where $\mathbf{\Pi}$ is a $K\times K$ permutation matrix, and $\mathcal{P}_{K}$ denotes the set of all such permutation matrices. Fig.~\ref{fig:rmse} shows the RMSE performance versus SNR. In the low-SNR regime from $-5$ dB to $5$ dB, all four methods exhibit a threshold effect because noise dominates the single-snapshot observation. As the SNR increases beyond $10$ dB, square Hankel MUSIC and truncated Hankel MUSIC with dense grid search reach an error floor of approximately $-33.6\,\mathrm{dBrad}$. This floor is caused by the finite angular grid spacing. In contrast, the Newton-refined methods reduce the grid-induced quantization error and continue to improve in the high-SNR regime. The performance gap between truncated Hankel Newton-MUSIC and square Hankel Newton-MUSIC remains within approximately $1.5$ dB over the simulated SNR range, indicating that the proposed truncation substantially reduces complexity while preserving most of the estimation accuracy.

\begin{figure}[t]
\centering
\includegraphics[width=0.88\linewidth]{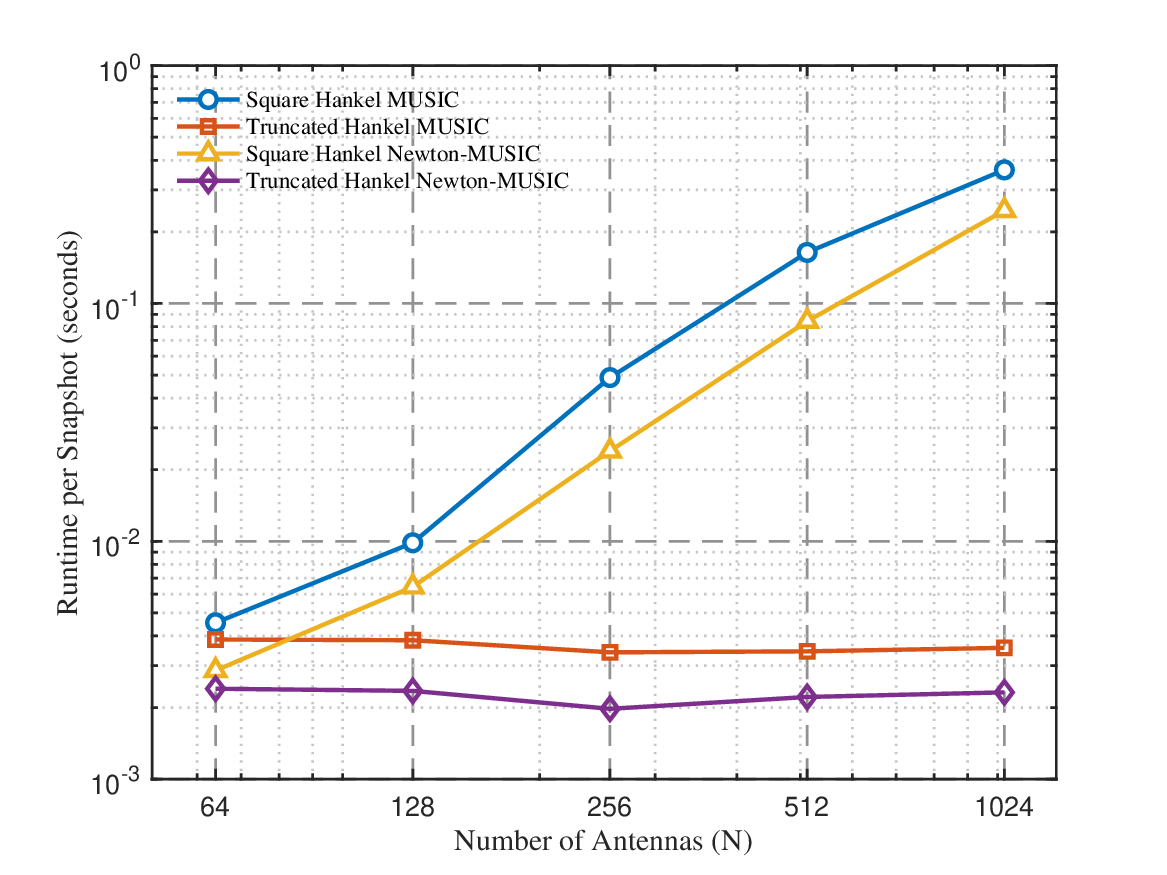}
\caption{Average runtime per snapshot versus number of antennas.}
\label{fig:time}
\end{figure}

Fig. \ref{fig:time} shows the average runtime per snapshot versus the number of antenna ports $N$ on a log-log scale. In this simulation, square Hankel methods use $L_{\mathrm{O}}=N/2$, whereas truncated Hankel methods use $L_{\mathrm{N}}=20$, and all other parameters remain unchanged. As $N$ increases from $64$ to $1024$, the runtime of square Hankel methods grows rapidly due to the cubic cost of high-dimensional correlation construction and subspace decomposition. In contrast, the truncated Hankel methods show a much slower growth rate because the row dimension of the smoothed matrix is fixed. The proposed truncated Hankel Newton-MUSIC method achieves the lowest average runtime among the compared methods. This reduction is mainly due to the use of a coarse grid for initialization and a small number of Newton iterations for off-grid refinement. At $N=1024$, the proposed method achieves more than two orders of magnitude runtime reduction compared with conventional square Hankel MUSIC, which is consistent with the linear scaling predicted by the complexity analysis.

\section{Conclusion}
This paper presented a low-complexity Truncated Hankel Newton-MUSIC framework for single-snapshot DoA estimation in reconfigurable antenna-enabled ISAC systems. By constructing a truncated Hankel matrix with a fixed row dimension, the proposed method reduces the dimension of the smoothed correlation matrix and lowers the cost of subspace decomposition. When the truncation length is fixed independently of the array size, the dominant complexity scales linearly with the number of antenna ports. A second-order Newton refinement step was further introduced to refine coarse grid estimates in the continuous angular domain, thereby reducing grid-induced quantization errors. Simulation results showed that the proposed method achieves accuracy close to square Hankel Newton-MUSIC while reducing the runtime by more than two orders of magnitude for large arrays. These results indicate that the proposed framework is a promising candidate for real-time DoA estimation in dynamic ISAC scenarios.

\bibliographystyle{IEEEtran}
\bibliography{reference.bib}

\end{document}